\pdfoutput=1
\documentclass[10pt,a4paper]{article}
\RequirePackage{latexsym,amsmath,amssymb}
\RequirePackage[dvipsnames,usenames]{color}

\usepackage{fullpage}
\usepackage{cancel}
\usepackage[english]{babel}
\usepackage[T1]{fontenc}
\usepackage[final]{showkeys} 
\usepackage{hyperref}
\hypersetup{pdfauthor={M. Cadoni, A. Sanna...},
            pdftitle={}
            }
\usepackage[]{cleveref}
\Crefname{equation}{Eq.~}{Eqs.}
\Crefname{figure}{Fig.}{Figs.}
\crefformat{plural}{#2eqs.~(#1)#3}
\crefname{section}{Sect.}{Sects.}

\usepackage{amsfonts}
\usepackage{graphicx}
\def\lb{\label}
\def\be#1\ee{\begin{align}#1\end{align}}

\def\a{\alpha}
\def\b{\beta}

\def\lb{\label}

\title{\bf  Emergence of a Cosmological Constant in Anisotropic Fluid Cosmology}
\author{M.~Cadoni${}^{ab}$\thanks{E-mail: mariano.cadoni@ca.infn.it},
A. P. ~Sanna$^{ab}$\thanks{E-mail: asanna@dsf.unica.it} \ 
\\
\\
${}^a$\emph{Dipartimento di Fisica, Universit\`a di Cagliari}
\\
{\em Cittadella Universitaria, 09042 Monserrato, Italy}
\\
\\
${}^b$\emph{I.N.F.N, Sezione di Cagliari, Cittadella Universitaria, 09042 Monserrato, Italy}
\\
\\}
\begin{document}
\maketitle
\begin{abstract}
We investigate anisotropic fluid cosmology in a situation where the spacetime metric  back-reacts  in a local,   time-dependent  way to the presence  of inhomogeneities. We derive exact solutions to  the Einstein field equations  describing   Friedmann-Lemaitre-Robertson-Walker (FLRW) large scale cosmological evolution in the presence of local inhomogeneities  and  time-dependent back reaction.  We use our derivation to tackle the cosmological constant problem.  A cosmological constant emerges by averaging the back reaction term on spatial scales of the order of $100 \ \text{Mpc}$, at which our universe begins to appear homogeneous and isotropic.  We find that the order of magnitude of the ``emerged'' cosmological constant agrees with astrophysical observations and is related in a natural way to baryonic matter density. Thus, there is no coincidence problem in our framework.

\end{abstract}
\section{Introduction}
\lb{sec1}
An anisotropic fluid  is a two-fluid model, which   can be used as an effective description of the sources of the gravitational field in several   regimes of the gravitational interaction \cite{Cosenza1981,Herrera1997,Cadoni:2017evg,Cadoni:2020izk, Herrera:2020gdg}.  Originally proposed to describe local anisotropies in  self-gravitating systems \cite{Cosenza1981,Herrera1997,Herrera:2020gdg},  recently it has gained renewed interest  in the emergent and corpuscular \cite{Cadoni:2017evg,Cadoni:2018dnd,Tuveri:2019zor, Verlinde:2016toy, Dvali:2011aa, Casadio:2020nns, Cadoni:2020mgb}  gravity  context.  In this framework it has been used to give a unified description  of baryonic matter, dark energy and their possible interaction \cite{Cadoni:2017evg}.  Specifically, at galactic scales, it can be used to explain the phenomenology commonly attributed to dark matter, in terms of a dark force originated by the reaction of dark energy (dark energy condensate in the corpuscular description) to the presence of baryonic matter. The  corresponding additional acceleration component is generated by the radial pressure of the anisotropic fluid  \cite{Cadoni:2017evg}.         

In a recent paper, the anisotropic fluid approach has been extended to the  cosmological regime  \cite{Cadoni:2020izk}. It has been shown that  this type of fluid can be used to describe the transition from an inhomogeneous universe at short scales to a homogeneous one at large scales, during the dark energy-dominated epoch. The anisotropic stress of the fluid  generates inhomogeneities and  therefore could be responsible for structure formation at small scales. Using cosmological perturbation theory, a  power spectrum for mass density distribution behaving as $1/k^4$ was found, in good accordance with  the observed two-point correlation function for matter distribution at small scales \cite{Cadoni:2020izk}.

The results  of Ref. \cite{Cadoni:2020izk} were obtained in the decoupling limit, where the FLRW cosmological dynamics for the scale factor  fully decouples from that describing inhomogeneities.  This decoupling of degrees of freedom occurs both at the level of  exact field equations and cosmological perturbation. The use of the decoupling limit is fully justified as long as  one is only interested  in structure formation at small scales in the dark energy-dominated era. On the other hand, one generically expects   the presence of inhomogeneities to generate 
a time-dependent back reaction of the geometry.

Inclusion of the   back reaction could be relevant for the  cosmological model, i.e. the dynamics of the scale factor $a$. In particular, it could be in principle related to the appearance  of a  cosmological  constant. 

This is a quite interesting possibility  because it could allow us to tackle the cosmological constant  and the coincidence problems \cite{Casadio:2020nns, Peebles:2002gy, Velten:2014nra, Copeland:2006wr}, i.e to explain why the cosmological constant has exactly the value we observe  and why in the present epoch it has the same  order of magnitude of  the density of baryonic matter.
Moreover, showing the existence of a  relation between  back reaction of the spacetime geometry and the presence of a cosmological constant  would corroborate the emergent gravity scenarios, where the latter is assumed, but not explained. \\
A related issue is the  physical interpretation of the cosmological constant. It can be regarded as either a purely geometric term, when written on the l.h.s. of Einstein equations  or as a source term, when written on the r.h.s. While  the standard  FLRW cosmology is based on the latter interpretation, the emergent gravity scenario  naturally selects the former. This is consistent with the standard back reaction approach, which interprets the observational evidence of the accelerated expansion of the universe not as the presence of an exotic source, but as a non-linear response of   the  spacetime geometry to cosmological matter perturbations. \\

Following this line of thoughts, in this paper we  derive, using an appropriate  parametrization of the spacetime metric,  exact solutions of Einstein's equations sourced by an anisotropic fluid.   The model describes a FLRW universe at large scales, local inhomogeneities at small scales together with  local, time-dependent back reaction of the geometry. We then apply  these results  to  understand the origin of the cosmological constant. 
It emerges when we     average the back reaction term on the spatial  scales  at which our universe appears homogeneous and isotropic.
The resulting cosmological constant 
 is related   in a natural way to   baryonic matter density - this  solves the coincidence problem -  and its order of magnitude   is in agreement with astrophysical observations. 

The structure of this paper is the following.
In section~\ref{sec2} we briefly review the basics of  anisotropic fluid cosmology.
 In section~\ref{sec3} we discuss the  cosmological dynamics, including the  back reaction of the metric to the presence of inhomogeneities.  In section~\ref{sec4} we explore the emergence of a cosmological  constant. Finally in section~\ref{sec5} we state our conclusions.
\section{Anisotropic fluid cosmology}
\lb{sec2}
Anisotropic fluid cosmology is a generalization  of the usual FLRW cosmology, which uses    an anisotropic fluid to source  cosmological evolution instead of the perfect fluid with barotropic equation of state.
It is a two-fluid model, whose stress-energy tensor,  after an appropriate redefinition  of the  velocities of the two barotropic fluids,  can be written in terms of an anisotropic fluid with velocity $u_{\mu}$ and a spacelike vector $w_{\nu}$  (see, e.g. Ref. \cite{Bayin:1985cd}).  The stress-energy tensor  for the fluid takes the form 
\begin{equation}
T_{\mu\nu} = \left(\rho + p_{\perp} \right)u_{\mu}u_{\nu} + p_{\perp} \ g_{\mu\nu} - \left(p_{\perp}-p_{\parallel} \right)w_{\mu}w_{\nu},
\label{TensoreEI}
\end{equation}
where $u_{\mu}$ and  $w_{\nu}$ satisfy  $g^{\mu\nu} u_{\mu}u_{\nu} = -1$, $g^{\mu\nu} w_{\mu}w_{\nu} = 1$, $u^{\mu}w_{\mu} = 0$ and  $\rho,\, p_{\parallel},\, p_{\perp}$ are respectively  the fluid  density,  pressures parallel  and perpendicular  to the spacelike vector $w_{\nu}$  \cite{Cosenza1981,Herrera1997}.

Taking  $p_{\perp}= p_{\parallel}\equiv p$ and assuming  a spatially homogeneous and isotropic spacetime metric,  we get the usual FLRW cosmological model with $p, \ \rho$ and the scale factor of the metric depending on the cosmological time  only. In this particular case, cosmological evolution is driven by  a perfect fluid with  barotropic  equation of state $p=p(\rho)$, whereas the velocity field $u^{\nu}$ is free from rotation,  shear and acceleration.

A non trivial anisotropic fluid  is obtained   by taking $p_{\perp}\neq  p_{\parallel}$, generating an anisotropic stress.
An important feature is that, although  we have anisotropy  in the  stress-energy tensor, the spacetime metric may remain  isotropic if homogeneity of the geometry    is broken  by allowing $p_{\perp}, \ p_{\parallel}$ and $\rho$ to develop a dependence from the radial coordinate $r$.
These features of the anisotropic fluid have been used  (a) in Ref. \cite{Cadoni:2017evg} to generate through $ p_{\parallel}(r)$ a "dark force" at galactic level,  which can explain  the galactic phenomenology without assuming the existence of dark matter, and (b) in Ref. \cite{Cadoni:2020izk} in
the cosmological framework, to describe  structure formation at small scales in the late, dark-energy dominated universe. On the other hand, one should keep in mind that the anisotropic  fluid   cosmological model is  unable   to describe the early-time cosmology and the radiation/baryonic matter dominated eras, at least in the form given in  Ref. \cite{Cadoni:2020izk}.  This is because any mass distribution is intrinsically unstable in the FLRW cosmology. \\

In the setup of Ref. \cite{Cadoni:2020izk}, the spacetime metric is assumed  to be isotropic  but not homogeneous. It  turns out that it is determined by two metric functions only  - the conformal factor  $a(t)$ and a metric function $f(r)$- which separately depend on the conformal time $t$ and the radial coordinate $r$:
\begin{equation}
ds^2 = a^2(t)\left[-dt^2 + f^{-1} (r) dr^2 + r^2 d\Omega^2\right]; \hspace{0.5 cm} d\Omega^2 = d\theta^2 + \sin^2 \theta \ d\phi^2.
\label{metricaold}
\end{equation}
The scale factor  $a(t)$ describes the cosmological dynamics, whereas $f(r)$ is determined by the inhomogeneities   in the distribution of baryonic matter at small scales. 
A  nice feature of the parametrization~(\ref{metricaold}) is that the dynamics for  $a(t)$ and $f(r)$ fully decouples \cite{Cadoni:2020izk}. 
The  scale factor $a$ is determined by FLRW cosmological equations, whereas $f(r)$ is determined by the  
the Misner-Sharp mass $m_B(r)$ associated with the inhomogeneity density $\mathcal{E}(r)$  in the baryonic matter,
 $m_B(r)=4\pi  \int dr \ r^2 \mathcal{E}(r)$, through the equation \cite{Cadoni:2020izk}:
\be \lb{g1a}
f= 1- \frac{2 G m_B(r)}{r}.
\ee
The inhomogeneities generate  a contribution $\mathcal{P}(r)$ to the radial pressure $p_{\parallel}$ given by \cite{Cadoni:2020izk}:
  \be\lb{press}
  \mathcal{P}(r)= -\frac{m_B(r)}{4\pi r^3}.
  \ee
The  decoupling  between $a$ and $f$ is  consequence of the parametrization of the metric~(\ref{metricaold}), which neatly separates time dependence from radial dependence.  This is   natural and satisfactory as long as we are interested in  describing cosmological dynamics at large scales and structure formation at small scales in a dark energy-dominated era. However,  in a generic situation, we expect  a time-dependent back reaction of the metric to the presence of local inhomogeneities. In this situation, the metric parametrization~(\ref{metricaold}) is not satisfactory anymore and we expect the dynamics of $a$ to be coupled to that of $f$. This will be the topic of the  next section.

\section{Inhomogeneous cosmological dynamics with  back reaction on the metric}
\lb{sec3}
In order to describe   time-dependent back reaction of the metric to the presence  of inhomogeneities in the context of anisotropic fluid cosmology, we need  a generalization of the metric parametrization~(\ref{metricaold}).   We want to preserve spherical symmetry, we will therefore assume the following: 
\begin{equation}
ds^2 = a^2(t)\left[-e^{\alpha(t, r)} dt^2 + e^{\beta(t, r)} dr^2 + r^2 d\Omega^2\right],
\label{metrica}
\end{equation}
where  $\alpha(t, r)$ and $\beta(t, r)$ are metric functions depending on both  $t$ and $r$.

The resulting Einstein equations, sourced by an anisotropic fluid with stress-energy tensor given by 
Eq.~(\ref{TensoreEI}),
 take the form:
 \begin{equation}
3\frac{\dot{a}^2}{a^2}+\frac{e^{\alpha-\beta}}{r^2}\left(-1+e^{\beta}+r\beta' \right)+\frac{\dot{a}}{a}\dot{\beta} = 8\pi G e^{\alpha} a^2 \ \rho;
\label{E00}
\end{equation}
\begin{equation}
\frac{\dot{a}}{a}\alpha'+\frac{\dot{\beta}}{r}=0;
\label{E0r}
\end{equation}
\begin{equation}
\frac{\dot{a}^2}{a^2}e^{\beta-\alpha}+\frac{1-e^{\beta}+r\alpha'}{r^2}+e^{\beta-\alpha} \left(-2\frac{\ddot{a}}{a}+\frac{\dot{a}}{a}\dot{\alpha} \right)=8\pi G e^{\beta} a^2 p_{\parallel};
\label{Err}
\end{equation}
\begin{equation}\begin{split}
&\frac{e^{-\left(\alpha+\beta\right)}}{4a^2}r \biggl\{4e^{\beta} r\dot{a}^2+4e^{\beta}r\ a\dot{a}\left(\dot{\alpha}-\dot{\beta} \right)+a \biggl[-8e^{\beta} r \ddot{a}+a \biggl[e^{\alpha}r \alpha'^2-2e^{\alpha}\beta'-e^{\alpha} \alpha' \left(-2+r\beta' \right) +2e^{\alpha} r \alpha''+\\
&+e^{\beta} r \left(\dot{\alpha}\dot{\beta}-\dot{\beta}^2-2\ddot{\beta} \right)\biggr] \biggr] \biggr\}=8\pi G a^2 r^2 \ p_{\perp}.
\label{Ethethe}
\end{split}\end{equation}
 The  dot  and the prime denote derivatives with respect to $t$ and $r$, respectively.

Conservation of the stress-energy tensor gives two more equations:
\begin{equation}
\dot{\rho}+\frac{\dot{a}}{a}\left(3\rho+p_{\parallel}+2p_{\perp} \right)+\frac{\dot{\beta}}{2}\left(\rho+p_{\parallel}\right)=0,
\label{T0}
\end{equation}
\begin{equation}
p'_{\parallel}+\frac{\alpha'}{2}\left(\rho+p_{\parallel}\right)+\frac{2}{r}\left(p_{\parallel}-p_{\perp}\right)=0.
\label{Tr}
\end{equation}
Eq.~(\ref{E0r}) can be immediately integrated to give
\begin{equation}
\alpha = \mathcal{A}(t) -\frac{a}{\dot{a}}\int \dot{\beta}\frac{dr}{r},
\label{alphabeta}
\end{equation}
where $\mathcal{A}(t)$ is an arbitrary integration function. 

We can now  use Eq.~(\ref{Tr}) to express $p_{\perp}$ in terms of $p_{\parallel}$ and $\rho$:
\begin{equation}
p_{\perp} = \frac{r}{2}p'_{\parallel}+p_{\parallel}+\frac{\alpha'\ r}{4}\left(\rho+p_{\parallel}\right).
\label{pperp}
\end{equation}

Using this result, the field equations~(\ref{E00}), (\ref{Err}) and (\ref{Ethethe}) can be appropriately combined to give:
 \begin{equation}
 \label{eq1}
 \dot{\alpha}\dot{\beta} =0.
\end{equation}

It follows that the field equations allow for two classes of solutions characterized  either  by $\dot{\beta}=0$ 
or by  $\dot{\alpha}=0$. Let us discuss the two cases separately.

\subsection{The decoupling limit}
\lb{subsec3}

The case  $\dot{\beta}=0$  corresponds to the decoupling  of   cosmological dynamics (the dynamics of the scale factor $a$)   from that of inhomogeneities   discussed in Sect.~\ref{sec2}. 
In fact,  in this case  the metric function $e^{\beta}$ is a function of the radial coordinate $r$ only. Moreover, Eq.~(\ref{alphabeta}) implies $\alpha = \mathcal{A}(t)$ so that  the metric function $e^{\alpha}$  can be completely absorbed by a redefinition of the conformal time $t$.  We are left  with only two degrees of freedom, the conformal factor of the metric $a$,  which describes the cosmological dynamics, and the metric function $e^{\beta(r)}$, which describes inhomogeneities. Using the  notation of Eq.~(\ref{metricaold})  we have $e^{\beta(r)}=f^{-1}$,  with $f$ given by Eq.~(\ref{g1a}).

In this situation we have exactly the case discussed   in Sect.~\ref{sec2}, where we have seen that the  cosmological dynamics  for $a$  is described by FLRW equations and decouples completely from that of inhomogeneities described by $e^{\beta(r)}=f^{-1}$. 
Notice that, owing to Eq.~(\ref{alphabeta}), the case $\dot{\beta}=0$  is completely equivalent to $\dot{\beta}=\dot{\alpha}=0$.

\subsection{Cosmological dynamics with back reaction of the metric}

The case $\dot{\alpha}=0$, when the metric function $e^{\a}$ depends only on $r$, describes the situation in which the metric responds  to the presence of inhomogeneities by developing   also time dependence of the metric function $e^{\beta}$. The cosmological degree of freedom $a$ is not decoupled from $\a$ and $\b$ anymore. The dynamical equation for $a$ will therefore depend on the presence of the inhomogeneities (back reaction).

Being $\a=\a(r)$ a function of the radial coordinate  $r$ only, solving Eq.~(\ref{alphabeta})  for $\b$ yields
\begin{equation}\lb{solbeta}
 e^{-\beta} =g(r) a^{r\a'(r)}, 
\end{equation}
where $g(r)$ is an  integration function.

Using Eq.~(\ref{E0r}), the Einstein field equations and energy-momentum conservations take the form 
\begin{subequations}
\begin{align}
&\frac{\dot{a}^2}{a^2}\left(3-r\alpha' \right)e^{-\alpha} + \frac{1-e^{-\beta}+r\beta'e^{-\beta}}{r^2} = 8\pi Ga^2 \rho; \label{alpha00}\\
&\frac{e^{-\beta}+re^{-\beta} 	\alpha'-1}{r^2}+e^{-\alpha}\left(-2\frac{\ddot{a}}{a}+\frac{\dot{a}^2}{a^2} \right)=8\pi G a^2 p_{\parallel} ; \label{alpharr}\\
&\dot{\rho} + \frac{\dot{a}}{a}\left(3\rho+3p_{\parallel}+rp'_{\parallel} \right)=0, \label{alphaconserv}
\end{align}
\end{subequations}
together with Eq.~(\ref{solbeta}) giving $\beta$ in terms of $\a$ and  the arbitrary function $g(r)$.
As usual in FLRW cosmology, the conservation equation~(\ref{alphaconserv}) is not independent, but is a consequence of the field equations, so that we are left with a system of three independent equations for $\a(r)$, $\b(r,t)$, $a(t)$, $\rho(r,t)$, $p_{\parallel}(r,t)$.

General,  analytic solution of Eqs.~(\ref{solbeta}), (\ref{alpha00}), (\ref{alpharr}) are difficult to find. Following a method similar to that used in Ref. \cite{Cadoni:2020izk}, we will look for solutions by separating the equations  for  the conformal factor $a$ from those for  $\a$ and $\b$.  Although a full decoupling of these degrees of freedom is not possible in this case, one can still rewrite  Eqs.~(\ref{alpha00}), (\ref{alpharr}) in terms of separate  equations. This can be very useful when the time-scale  of variation  of $a$ is much larger than that of the function $\b$.

Let us first define  rescaled  density $\hat{\rho}$ and pressure  $\hat{p}_{\parallel}$ for the fluid:
\begin{equation}\lb{eq6}
\hat{\rho} \equiv \frac{3e^{\alpha}}{3-r\alpha'}\rho; \qquad \hat{p}_{\parallel}\equiv e^{\alpha} p_{\parallel} .
\end{equation}
Rewritten in terms of these rescaled quantities, Eqs.~(\ref{alpha00}), (\ref{alpharr}) take the form:
\begin{subequations}
\begin{align}
&a^2 \hat{\rho} =  \frac{3}{8\pi G} \left(\frac{\dot{a}}{a} \right)^2+\frac{3 e^{\alpha}}{8\pi G r^2}\frac{\partial_r \left(r-re^{-\beta} \right)}{3-r\alpha'} \label{alpha004};\\
&a^2 \hat{p}_{\parallel}=\frac{1}{8\pi G}\left[\left(\frac{\dot{a}}{a} \right)^2-2\frac{\ddot{a}}{a}\right] +\frac{e^{\alpha}}{8\pi G r^2} \left(e^{-\beta}+re^{-\beta}\alpha'-1\right). \label{alpharr4}
\end{align}
\end{subequations}

Let us now assume that $\hat{\rho}$ and $\hat{p}_{\parallel}$ can be separated in $(1)$ a homogeneous, time-dependent part,  $(2)$ a inhomogeneous time-independent part and $(3)$ a $(t,r)$-dependent  interaction part:
\begin{subequations}
\begin{align}
&a^2 \hat{\rho}(t,r) \equiv  a^2{\rho}^{(1)}(t)+\frac{3 e^{\alpha}}{3-r\alpha'}\left(\rho^{(2)}(r)+ \rho^{(3)}(r,t)\right), \label{a2rho}\\
&a^2 \hat{p}_{\parallel}(t,r) \equiv  a^2{p}_{\parallel}^{(1)}(t)+ e^{\alpha}\left({p}_{\parallel}^{(2)}(r)+ {p}_{\parallel}^{(3)}(r,t)\right).
\end{align}
\label{hatquatities}
\end{subequations}
The physical intuition  behind  the ansatz~(\ref{hatquatities}) is that the anisotropic fluid sourcing gravity  allows for a separation between the  purely time-dependent cosmological dynamics  sourced by ${\rho}^{(1)}(t)$ and that of inhomogeneities sourced by ${\rho}^{(2)}(r)$, whereas $\rho^{(3)}(r,t)$ represents an interaction  term (back reaction) whose form we expect  to be  constrained by the sources ${\rho}^{(1)}$  and ${\rho}^{(2)}$. 

Using the ansatz~(\ref{hatquatities}) we can separate  Eqs.~(\ref{alpha004}), (\ref{alpharr4}) into $(1)$ a purely time-dependent part, which determines the scale factor $a$ and takes a    FLRW form; $(2)$ a $(t,r)$ part, which determines the metric function $\beta$ in terms of the densities $\rho^{(2,3)}$, and  $(3)$ a relation determining the pressures   ${p}_{\parallel}^{(2,3)}$.

More specifically we get from the purely time-dependent part
\begin{subequations}
\begin{align}
&a^2{\rho}^{(1)}(t) = \frac{3}{8\pi G}\left(\frac{\dot{a}}{a} \right)^2, \label{tilderho}\\
&a^2 {p}_{\parallel}^{(1)}(t) =\frac{1}{8\pi G}\left[\left(\frac{\dot{a}}{a} \right)^2-2\frac{\ddot{a}}{a} \right], \label{tildep}
\end{align}
\end{subequations}
which are the FLRW cosmological equations written in conformal frame $ds^2 = a^2(t)(- dt^2 +  dr^2 + r^2 d\Omega^2)$ for  the case of zero spatial curvature.

The metric function $e^{-\beta}$ can be determined from  Eq.~(\ref{alpha004}) and turns out to be:
\be\lb{eq6}
e^{-\b}= 1-\frac{2 G M(t,r)}{r},\quad  M(t,r)\equiv M^{(1)}(t)+M^{(2)}(r)+M^{(3)}(r,t).
\ee
Here the time-dependent mass term $M^{(1)}(t)$ is an integration function, whereas $M^{(2,3)}$ are the masses associated with inhomogeneities  and the interaction,  respectively:
\be\lb{eq8}
M^{(2)}(r)=4\pi\int dr\, r^2 \rho^{(2)}(r),\quad M^{(3)}(r,t)=4\pi\int dr \,r^2 \rho^{(3)}(r, t).
\ee

The other metric function $\a$  is then determined using Eq.~(\ref{alphabeta}), giving:
\be\lb{eq9}
\alpha =\frac{1}{\ln a} \int \frac{dr}{r}\ln \left(\frac{1}{g}-\frac{2GM}{gr} \right).
\ee

Eq.~(\ref{alpharr4}) determines $\beta$  in terms of  ${p}_{\parallel}^{(2,3)}$ and $\a$, which upon using Eqs.~(\ref{eq6}), (\ref{eq9}) gives:
\be\lb{eq10}
{p}_{\parallel}^{(2)}=-\frac{M^{(2)}(r)}{4\pi r^3},\quad {p}_{\parallel}^{(3)}= -\frac{M^{(1)}(t)+M^{(3)}(r, t)}{4\pi r^3} +\frac{1}{8\pi G r^2 \ \ln a} \left(1-\frac{2GM}{r} \right)\ln\left(\frac{1}{g}-\frac{2GM}{gr^2} \right).
\ee

Notice that the solutions~(\ref{eq6}), (\ref{eq8}), ${p}_{\parallel}^{(2)}$ and the first term in ${p}_{\parallel}^{(3)}$ of~(\ref{eq10}) have the same form of those appearing in Eqs.~(\ref{g1a}) and (\ref{press})  in  the decoupling limit.
The mass  $M(r,t)$ in Eq.~(\ref{eq6}) is a $t,r$-dependent Misner-Sharp mass. Moreover, the  decoupling limit discussed in Sect.~\ref{sec2}   is obtained in the limit  $\a=\text{const}$.
In this limit we can easily see that the interaction term is switched off: $M^{(1)}=M^{(3)}={p}_{\parallel}^{(3)}=0$ so that   equations~(\ref{eq6}), (\ref{eq8}) match exactly Eqs.~(\ref{g1a}) and  (\ref{press}).

It is important to stress that Eq.~(\ref{eq9})   implies a  strong constraint on the form of the mass function $M(t,r)$. This is because the l.h.s of Eq.~(\ref{eq9}) is a function of the radial coordinate $r$ only, whereas in its r.h.s functions of both $t$ and $r$ are present. $M$ has to be chosen in such way to cancel the overall dependence on $t$.  As anticipated, this constraint  is largely expected from a physical point of view. The only completely free contributions to~(\ref{a2rho}) are the spatially homogeneous $\rho^{(1)}$ and inhomogeneous $\rho^{(2)}$ densities. They are determined  respectively by the cosmological evolution and by the physics of the early universe,  for which our cosmological model  cannot be    used  as stated in  Sect.~\ref{sec2}. Their interaction term $\rho^{(3)}$, on the other hand, is  not completely free but is constrained by the  form of $\rho^{(1)}$ and $\rho^{(2)}$.  A way to encode information about  early time cosmology  in  our context is to assume a phenomenological relation   between ${p}_{\parallel}^{(2)} $  and  $\rho^{(2)}$, which would  play a role analogous to the equation of state for the fluid  in the usual FLRW cosmology. In Ref.   \cite{Cadoni:2020izk}  such  relation was inspired by galactic dynamics.  With the same logic one  could assume a relation between ${p}_{\parallel}^{(3)} $  and  $\rho^{(3)}$, but we will not elaborate on this point in this paper. \\

Finally,  once $p_{\parallel}$ and $\rho$ are known,  the perpendicular pressure of the fluid can be easily  computed using Eq.~(\ref{pperp}). 

\section{Emergence of the cosmological constant}
\lb{sec4}
In  the emergent gravity  framework, geometry and gravity are seen as macroscopic features emerging out of some  quantum mechanical microscopic theory.  The phenomenology  commonly attributed to dark matter could also be explained  as an emergent phenomenon,   a dark force  originated from the response of  dark  energy to the presence of baryonic matter \cite{Verlinde:2016toy}. 
An interesting realization  of this scenario has been  given in the  corpuscular  gravity context  \cite{Cadoni:2017evg, Cadoni:2018dnd, Tuveri:2019zor},  where an effective description of dark energy and  baryonic matter in terms of an anisotropic fluid has been proposed.  

In this emergent gravity framework, dark energy is assumed to be present from the beginning and until now no attempt to explain its origin has been made. On the other hand, the results of Ref. \cite{Cadoni:2020izk} show that also matter distribution at small cosmological  scales can be explained in anisotropic fluid cosmology as due to inhomogeneities  triggered by the fluid anisotropy. The results of  Ref. \cite{Cadoni:2020izk} have been derived in the regime  of cosmological dynamics  for the scale factor $a$ fully decoupled from that of the inhomogeneities. It is  therefore tempting to  look  at   the  interaction term in    Eqs.~(\ref{alpha00}), (\ref{alpharr}), which describes the reaction of the geometry to the presence of inhomogeneities,   as the source for dark energy.

In order to keep the discussion as simple as possible, we will consider  the simplest case of dark energy, namely that of a cosmological constant $\Lambda$.  In  presence of a $\Lambda$ term, Eqs.~(\ref{tilderho}) and (\ref{tildep}) become:

\begin{subequations}
\begin{align}
&a^2{\rho}(t) = \frac{3}{8\pi G}\left(\frac{\dot{a}}{a} \right)^2-\frac{\Lambda}{8\pi G} a^2; \label{tilderho1}\\
&a^2 {p}(t) =\frac{1}{8\pi G}\left[\left(\frac{\dot{a}}{a} \right)^2-2\frac{\ddot{a}}{a} \right] +\frac{\Lambda}{8\pi G} a^2. \label{tildep1}
\end{align}
\end{subequations}

If we want to obtain  these  equations  from Eqs.~(\ref{alpha004}), (\ref{alpharr4})   we have to first solve two main conceptual difficulties.  First,  the general solution we found in Sect.~\ref{sec3} gives  interaction terms in  Eqs.~(\ref{alpha00}), (\ref{alpharr}), which are a complicate function  of the scale factor $a$. Moreover, as we have seen in the previous section,  Eq.~(\ref{eq9}) represents an  highly non trivial constraint on the  form of $\beta$.   Second, these terms  have not the form  of a constant multiplying the scale factor  $a^2$.

The first difficulty can be solved  by looking at Eqs.~(\ref{tilderho1}), (\ref{tildep1}) as the leading order  in the  expansion of 
Eqs.~(\ref{alpha004}), (\ref{alpharr4}) near the decoupling limit, i.e.  $\a=\text{constant}$ (or equivalently $r\a'=0$) and near the present evolution stage of our universe, i.e   $a^2=1$.  This second limit is also justified by the fact that, according to present observations \cite{Aghanim:2018eyx}, the cosmological constant starts dynamically dominating over matter in the very late universe, at $z \sim 0.6$.  \\
For what concerns the second difficulty, we will simply assume that  the cosmological term results from spatial averaging the interaction terms in Eqs.~(\ref{alpha004}), (\ref{alpharr4}) at scales $R\approx 100 \ \text{Mpc}$  at which our universe appears to be  homogeneous.

The use of spatially averaged quantities is well motivated  in cosmology.  The cosmological principle, i.e. the homogeneity and isotropy of the universe, is  valid only in a statistical sense and  at large scales, whereas our universe is highly non homogeneous at small scales. Moreover,   due to the non-linear character of Einstein's equations, spatial averaging and time evolution do not commute,  an aspect whose importance for cosmology was first investigated in \cite{Shirokov1998} and further emphasized in \cite{Ellis:1984bqf}. This implies that the field equations obtained from a smoothed metric (the FLRW equations in  the standard approach) are different from those obtained from a general inhomogeneous metric and then averaged on some scale. In the latter  case, new terms appear, which  describe the back reaction of inhomogeneities to geometry \cite{Buchert:1999er, Buchert:2001sa, Buchert:2011sx}. These terms  could potentially be important in the evolution of the universe (see, e.g. Refs. \cite{Buchert:2011sx, Ellis:2011hk}), for instance they could trigger the observed accelerated expansion of the universe \cite{Barausse:2005nf, Kolb:2005da, Buchert:2007ik, Celerier:2007jc,  Rasanen:2003fy, Rasanen:2006kp,Buchert:2002ij}.   

Expanding~(\ref{solbeta}) near $a^2=1$ and keeping only the  terms up to order $(a^2-1)$  we get:
\begin{equation}
\lb{exp1}
e^{-\b}\approx g(r) \left[ 1+ \frac{r\a'}{2}\left(a^2-1\right)\right].
\end{equation}

Using this  expression and the separation ansatz~(\ref{a2rho}) into Eq.~(\ref{alpha004}) and  separately equating terms depending only on $r$ and those containing also a $t$-dependence, we can find both $g$ and $\a'$: 
\be\lb{eq11}
g(r)=1- \frac{2G(M^{(2)}(r)+M^{(3)}(r))}{r},\quad g(r)\, r^2\a'=-4 G M^{(3)}(r)= -16\pi G\int dr \ r^2 \rho^{(3)}(r). 
\ee
Notice that  the compatibility constraint now requires $M^{(1)}(t)=0$ and $\rho^{(3)}$ to be a  function of  $r$ only.

Using back Eq.~(\ref{eq11}) into Eqs.~(\ref{alpha004}), (\ref{alpharr4}), keeping only the leading terms in the $\a=\text{constant}$ expansion we get:
\begin{subequations}
\begin{align}
&a^2 \hat{\rho} =  \frac{3}{8\pi G} \left(\frac{\dot{a}}{a} \right)^2+\rho^{(2)}(r)+ a^2 \rho^{(3)}(r)\label{alpha005};\\
&a^2 \hat{p}_{\parallel}=\frac{1}{8\pi G}\left[\left(\frac{\dot{a}}{a} \right)^2-2\frac{\ddot{a}}{a}\right] - \frac{M^{(2)}+2M^{(3)}}{4\pi r^3} - a^2 \frac{M^{(3)}}{4\pi r^3}.\label{alpharr5}
\end{align}
\end{subequations}

Eqs.~(\ref{alpha005}),  (\ref{alpharr5})  have been derived by starting from small cosmological scales, at  which the  decoupling limit discussed in Ref. \cite{Cadoni:2020izk} works. At this scales, inhomogeneities  determines the formation of structures encoded in the power spectrum for the density matter distribution \cite{Cadoni:2020izk}. We are here interested in the large scale cosmological dynamics.
The simplest way  to describe the transition to larger scales  is to  average Eqs.~(\ref{alpha005}) ,  (\ref{alpharr5})  over  the spatial coordinate $r$ at the typical scales at which our universe appears to be homogeneous and isotropic, i.e. $R \sim 100 \ \text{Mpc}$. 

The averaging procedure produces in    Eqs.~(\ref{alpha005}),  (\ref{alpharr5}) both constant terms and time-dependent terms.  One can easily see that the constant terms  on the l.h.s. of our equations cancel  those on  the r.h.s. Physically this cancellation  means that the time-independent part of the inhomogeneities plays a role only at small scales, where it is responsible  for structure formation (and for the emergence of a dark  force at galactic scales \cite{Cadoni:2017evg,  Tuveri:2019zor}),  whereas it is irrelevant at large scales. 
Moreover,  the cancellation of the constant terms stemming from averaging the $r$-dependent terms  in Eqs.~(\ref{alpha005}) and (\ref{alpharr5})  is consistent with the fact that the spatial curvature of our universe is zero \cite{Aghanim:2018eyx}.  In fact, a non-vanishing  constant term in  Eqs.~(\ref{alpha005}) and (\ref{alpharr5}) would give a non-zero spatial curvature. 

For what concerns  the  time-dependent part,  on the l.h.s of Eqs.~(\ref{alpha005}), (\ref{alpharr5}) we have the spatial averaging of 
$\hat{\rho}$, which contains  the terms depending on $\rho^{(1)}$ and $\rho^{(3)}$ of Eqs.~(\ref{hatquatities}) and similarly for 
$\hat{p}_{\parallel}$. Spatial averaging  of these quantities at large scales will produce the  observed  density $\rho(t)\equiv \langle  \hat{\rho} \rangle_r $ and pressure  $p(t)\equiv \langle  \hat{p}_{\parallel} \rangle_r $ of matter appearing in the resulting FLRW cosmological equations.

 After  taking the average over the spatial coordinate $r$ of~(\ref{alpha005}), (\ref{alpharr5})  we  get therefore the FLRW  cosmological equations:
\begin{align}
&a^2 \rho(t) =  \frac{3}{8\pi G} \left(\frac{\dot{a}}{a} \right)^2+ a^2 \langle\rho^{(3)}\rangle_r\label{alpha006};\\
&a^2 p(t) =\frac{1}{8\pi G}\left[\left(\frac{\dot{a}}{a} \right)^2-2\frac{\ddot{a}}{a}\right]
 - a^2 \Bigg\langle\frac{M^{(3)}}{4\pi r^3}\Bigg\rangle_r\label{alpharr6}.
\end{align}
In order to  calculate the $\rho^{(3)}$-depending  terms  in the previous equations, we need to make some educated guess about  the form of $\rho^{(3)}(r)$.   As $\rho^{(3)}$ is the energy density of the back reaction of the metric on the presence of the inhomogeneities, characterized by  $\rho^{(2)}$, these two have to be of the same order of magnitude, i.e. we have $\rho^{(3)}\sim \rho^{(2)}$ .  Moreover,  it is natural to assume that $\rho^{(3)}$ gives rise to an attractive force.  The fact that it has to behave as an inverse  power of $r$ implies  that  $\rho^{(3)}$ has to be  negative, as we will see in a while. 
The perturbative analysis of   Ref. \cite{Cadoni:2020izk} has shown that  $\rho^{(2)}\sim \kappa_1/r + \kappa/r^2$, where $\kappa,\kappa_1$ are some constants.
In  Ref. \cite{Cadoni:2020izk} it was also shown that $\kappa=0$ in the small scale regime. We are here considering the  large scale regime, at the transition to an homogeneous and isotropic universe,  in which inhomogeneities are expected to die out more rapidly than at small scales.  We are therefore led to put $\kappa_1=0$, so that we have:
\be\lb{eq23}
\rho^{(3)}\sim - \frac{\kappa}{r^2}.
\ee

 The behavior of $\rho^{(3)}\sim r^{-2}$  is consistent with the standard cosmological scenario and with  the results of Ref. \cite{Cadoni:2020izk}.
 The  large scale, inflationary, matter power spectrum  $P(k) \sim k$ (where $k$ represents the wave number),
 is modified, approximately at the equivalence epoch, by the transfer function  $T^2(k)$, i.e. $P(k) \sim k T^2(k)$. At small scales   we 
 have  $P(k) \sim k^{-3}$  (or also $P(k) \sim k^{-4}$  if one works with matter  correlation functions \cite{Cadoni:2020izk}).  The bending  scale is $L \sim 50 \ \Omega_0^{-1} \ h^{-2} \ \text{Mpc}\sim 300 \ \text{Mpc}$ \cite{Peebles:2002gy}. $L$ is therefore a bit higher than the  scale $R$ we are considering for our model. Therefore,  the $ \rho^{(3)}\sim r^{-2}$ behavior corresponds to the power spectrum $ P^{(3)}(k) \sim \frac{1}{k^2}$, which is intermediate between  the large scale behavior  $P(k) \sim k$  at scales bigger than  $300 \ \text{Mpc}$ and the small scale one $P^{(2)}(k) \sim  k^{-4}$.

Using Eq.~(\ref{eq23}) in Eq.~(\ref{alpha006}),(\ref{alpharr6}) we can easily see  that the equation  of state  $\langle{p}^{(3)}_{\parallel}\rangle_r= - \langle\rho^{(3)}\rangle_r$ holds. The same equations take the form given by Eqs.~(\ref{tilderho1}), (\ref{tildep1}). Hence, they describe  a  FLRW universe with  cosmological constant $\Lambda$ given by:
\be\lb{Lambdaestimate}
\Lambda\sim - 8\pi G\langle\rho^{(3)}\rangle_r\sim  8\pi G\langle\rho^{(2)}\rangle_r.
\ee
Hence, the cosmological constant term appears on the r.h.s. of  Eqs.~(\ref{alpha006}), (\ref{alpharr6}) as the spatial average of a term describing the back reaction of the geometry to the presence  of inhomogeneities.  The interpretation of the cosmological constant as a {\sl negative} term, i.e. a negative back reaction energy, in the in  r.h.s. of  Eqs.~(\ref{alpha006}) instead of a {\sl positive} term in its l.h.s., i.e a positive energy density of some form of matter, is fully consistent  with  the emergent gravity paradigm. The cosmological constant is not a form of (exotic) matter, but a  geometric term   describing  back reaction of the geometry to  the  presence of usual baryonic matter.

Let us now  estimate the order of magnitude of the cosmological constant given by Eq.~(\ref{Lambdaestimate}) and   compare our result with its  observed   value. Let us first write Eq.~(\ref{Lambdaestimate}) in terms of the cosmological parameter $\Omega_{\Lambda0}\equiv \Lambda/3\mathcal{H}_0^2$ and of the critical density $\rho_c = 3\mathcal{H}_0^2/8\pi G\sim 2 \cdot 10^{-29} \ \text{g} \cdot \text{cm}^{-3}$, where 
$\mathcal{H}_0$ is the Hubble parameter. We get: 
\be\lb{eq56}
\Omega_{\Lambda0}\sim \frac{\langle\rho^{(2)}\rangle_r}{\rho_c}.
\ee
The scale $R$ on  which we compute  spatial average of inhomogeneities is the transition scale at which our universe becomes homogeneous, this is about  $R \sim 100 \ \text{Mpc}$. $\rho^{(2)}(r)$ is the density profile of baryonic matter so that we have 
\be\lb{eq89}
\langle\rho^{(2)}\rangle_r= \frac{3M}{4\pi R},
\ee
where $M$ is the total baryonic mass inside a sphere of radius $R$. For $R \sim 100 \ \text{Mpc}$, we have $M \sim 10^{17} \ M_{\odot}$ and we get from  Eq.~(\ref{eq89})  $\Omega_{\Lambda0}\sim 10^{-1}$, which gives the correct order of magnitude of the observed cosmological constant \cite{Aghanim:2018eyx}.  Note that there is a strong dependence on the scale of averaging of the radial coordinate, namely our estimate agrees with observations only if we average at scales  $\sim 100 \ \text{Mpc}$.  The value of Eq.~(\ref{eq56}) changes drastically when changing the scale at which the average is performed.
The observed order of magnitude  of the cosmological constant is only obtained if the averaging scale is given by the  transition scale at which our universe begins to appear homogeneous and isotropic.
\\

Let us  conclude this section with a final remark. Our model explains the emergence of the cosmological constant as a  reaction of the geometry to inhomogeneities in the  baryonic matter density distribution. Therefore, there is no coincidence problem \cite{Peebles:2002gy, Velten:2014nra, Copeland:2006wr}. The fact that the energy densities associated to matter and $\Lambda$ are of the same order of magnitude at the present epoch is not a  coincidence, but it is linked to the intrinsic nature of the emerged cosmological constant.
Thus, in our model the  origin of the accelerated expansion of the universe can be traced back to structure  formation. Conversely,  the coincidence problem  represents a fierce challenge to the current standard cosmological model, since it requires extremely fine-tuned initial conditions in the early universe.  

\section{Conclusions}
\lb{sec5}

Building  on previous  work about anisotropic fluid cosmology \cite{Cadoni:2020izk},  in this paper we have found exact solutions of Einstein's field equations sourced by an anisotropic fluid, which  describe  FLRW cosmological evolution at large scales,  local inhomogeneities at small scales  and a local, time-dependent back reaction of the geometry to the presence of the latter.  These solutions generalize those  found in Ref. \cite{Cadoni:2020izk} to the case in which the FLRW cosmological evolution of the scale factor does not decouple completely from local inhomogeneities. They could find applications  as an alternative approach to the use of  perturbations in  late-time cosmology. In fact, in the usual setup of cosmological perturbation,  the background metric is   held fixed, whereas  our approach allows to compute the back reaction of the metric.  

A  simple way to compute the effect of back reaction  on the  cosmological dynamics   is to perform  spatial averaging of the inhomogeneous  terms.  This  is a quite conceptually involved point, because it reverses the usual cosmological paradigm, which first assumes homogeneity and isotropy  and then uses  perturbation theory in a homogenous and isotropic background. It is known that,  owing  to the non-linear character of Einstein's equations, spatial averaging and time evolution do not commute \cite{Shirokov1998, Ellis:1984bqf}. This implies that the field equations obtained  from a general inhomogeneous metric  averaged on some scale contain additional  terms, whose effect could potentially be important in the evolution of the universe. For instance,  they could trigger the observed accelerated expansion of the universe \cite{Barausse:2005nf, Kolb:2005da, Buchert:2007ik, Celerier:2007jc,  Rasanen:2003fy, Rasanen:2006kp,Buchert:2002ij}.

As an important application of this approach, we have considered the emergence of  a cosmological constant $\Lambda$ as the result of spatial averaging  the back reaction terms produced by local, time-dependent  inhomogeneities  in our anisotropic fluid cosmology.  Specifically, we were able to reproduce the  order of magnitude of the observed  cosmological constant by averaging on scales of order $100 \ \text{Mpc}$, at which our universe begins to appear homogeneous and isotropic.

Our solution of the cosmological constant  problem is  fully consistent  with the emergent gravity paradigm. In fact, $\Lambda$ is described  not as a form of (exotic) matter, but as a  geometric term  produced  by the   back reaction of the geometry to  the  presence of inhomogeneities  in the baryonic matter distribution. This means that  the accelerated expansion of our  universe can be traced back to structure formation  at  scales smaller than those  of the transition to homogeneity and isotropy.
Since   $\Lambda$ is determined by    the spatial average of baryonic matter,  the proposed solution to the cosmological constant problem has also the advantage of solving the  coincidence problem at the same time.
Being based on rough estimates of the parameters entering  in the model  (the mass $M$ and the averaging scale $R$), at least at the present stage of development, our model has not the predictive power to determine the  exact ratio between baryonic matter and   the  cosmological constant, but only orders of magnitude. 

Let  us conclude with some general remarks  about the use of anisotropic  fluids in cosmology.
They are very useful to describe  several regimes of gravity:  (1)  at galactic  scales, they produce an additional component  of the radial acceleration \cite{Cadoni:2017evg}; (2) at small scales in late-time cosmology, they can explain  structure formation \cite{Cadoni:2020izk}; (3) at large scale in late-time  cosmology, they may explain the emergence of a cosmological constant,  as shown in the present paper.  Their use in cosmology, however, has a strong limitation: they cannot be used  to describe early-time cosmology and the radiation/baryonic matter dominated eras.  Mass distribution is intrinsically unstable in FLRW cosmology. So, any inhomogeneous cosmological model can describe the late-time  epoch of the universe, cosmic structures at small scales  and the back reaction  of geometry  to the presence of the latter, but not the early-time one nor the evolution of perturbations  stemming from it.

This issue is strongly related with  the fact that we use inhomogeneities to generate  local anisotropic stresses -hence a non trivial anisotropic fluid. It may be possible that an alternative way to generate a non trivial  anisotropic fluid does exist, for instance  not  using inhomogeneities but  a different reparametrization of the original two-fluid description. Such an alternative description could open the way for using anisotropic  fluids also in early-time cosmology.

\end{document}